\documentclass[fleqn,usenatbib]{mnras}

\usepackage{newtxtext,newtxmath}

\usepackage[T1]{fontenc}

\DeclareRobustCommand{\VAN}[3]{#2}
\let\VANthebibliography\thebibliography
\def\thebibliography{\DeclareRobustCommand{\VAN}[3]{##3}\VANthebibliography}

\usepackage{graphicx}	
\usepackage{amsmath}	
\usepackage{amssymb}	
\usepackage{xcolor}
\usepackage{enumitem}
\usepackage{empheq}

\newcommand{\gae}{\lower 2pt \hbox{$\, \buildrel {\scriptstyle >}\over {\scriptstyle
\sim}\,$}}
\newcommand{\lae}{\lower 2pt \hbox{$\, \buildrel {\scriptstyle <}\over {\scriptstyle
\sim}\,$}}

\newcommand{\mjd}{\text{MJD}}

\newcommand{\dash}[1]{$-$ #1}

\newcommand{\pks}{PKS 2155--304}

\newcommand{\dgr}{$^\circ$}

\newcommand{\lf}{\left(}
\newcommand{\rt}{\right)}
\newcommand{\Bf}{\textit{B}}
\newcommand{\Vf}{\textit{V}}
\newcommand{\Rf}{\textit{R}}

\newcommand{\Jf}{\textit{J}}

\newcommand{\sy}{synchrotron}

\newcommand{\var}{\text{var}}

\newcommand{\cons}{\text{cons}}

\title[Photopolarimetry of \pks{} during a flare]{The optical polarization of the blazar \pks{} during an optical flare in 2010}

\author[N.~W. Peceur et al.]{N.\ W.\ Peceur$^{1}$\thanks{Contact e-mail: \href{mailto:nikki.pekeur@ast.uct.ac.za}{nikki.pekeur@ast.uct.ac.za}}, A.\ R.\ Taylor$^{1,2,3}$ and R.\ C.\ Kraan-Korteweg$^1$
\\
$^{1}$Department of Astronomy, University of Cape Town, Private Bag X3, Rondebosch 7701, SA
\\
$^{2}$Department of Physics and Astronomy, University of the Western Cape, Modderdam Road, Private Bag X17, Belville 7530, SA
\\
$^{3}$Inter-University Institute for Data Intensive Astronomy, University of Cape Town, Private Bag X3, Rondebosch 7701, SA}

\date{Accepted XXX. Received YYY; in original form ZZZ}

\pubyear{2020}

\begin{document}
\label{firstpage}
\pagerange{\pageref{firstpage}--\pageref{lastpage}}
\maketitle

\begin{abstract}
An analysis is presented of the optical polarimetric and multicolour photometric (\textit{BVRJ}) behaviour of the blazar PKS 2155--304 during an outburst in 2010. This flare develops over roughly 117 days, with a flux doubling time $\tau \sim 11$ days that increases from blue to red wavelengths. The polarization angle is initially aligned with the jet axis but rotates by roughly $90^\circ$ as the flare grows. Two distinct states are evident at low and high fluxes. Below 18 mJy, the polarization angle takes on a wide range of values, without any clear relation to the flux. In contrast, there is a positive correlation between the polarization angle and flux above 18 mJy. The polarization degree does not display a clear correlation with the flux. We find that the photopolarimetric behaviour for the high flux state can be attributed to a variable component with a steady power-law spectral energy distribution and high optical polarization degree (13.3\%). These properties are interpreted within the shock-in-jet model, which shows that the observed variability can be explained by a shock that is seen nearly edge-on. Some parameters derived for the relativistic jet within the shock-in-jet model are: $B=0.06$ G for the magnetic field, $\delta=22.3$ for the Doppler factor and $\Phi=2.6^\circ$ for the viewing angle.
\end{abstract}



\section{Introduction} \label{sec:intro}
Blazars are a subclass of radio-loud AGN, where the relativistic jet is closely aligned to the line-of-sight of the observer \citep{Urry1995} and for which the most extreme observational properties are detected \citep{Fan2000,HESS2007, Kastendieck2011}. The observed radiation spans the entire electromagnetic spectrum, from radio to $\gamma$--ray wavelengths, and is dominated by non-thermal emission from a relativistic jet, as well as the presence of polarization at radio and optical wavelengths. The spectral energy distribution (SED) consists of two broad emission features. The low-energy component is located at optical to soft $X$--ray energies and is due to relativistic electrons spiralling in a magnetic field (synchrotron emission). The high-energy component is located at hard $X$--ray to $\gamma$--ray energies and is attributed to Inverse Compton emission of the relativistic electrons. The seed photons for inverse Compton scattering can be provided by the synchrotron emitting electrons themselves (Synchrotron Self Compton emission), or from external photon fields (External Compton emission) \citep{HESS2009, Abdo2010b, Reynoso2012, Chen2012}. The magnetic field of the jet therefore underlies the physical processes that produce the observed blazar emission. 

The polarization is a direct observable of the magnetic field and can provide useful information on the geometry and degree of order of the magnetic field of the relativistic jet. The degree of optical polarization could be related to the level of ordering of the magnetic field or to the electron energy distribution within the emission region, while the position angle of the polarization vector could be related to the direction of the magnetic field vector along the line of sight \citep{Angel1980, Lister2000, Dulwich2009}.

Various models have been proposed to explain the observed polarization of blazars. These models can be divided into two classes: deterministic and stochastic. Stochastic models are based on lowering the maximum possible polarization degree for synchrotron radiation ($\sim 69\%$ for a power-law particle spectrum with index $p = 2$) to values that are more compatible with observations ($\lae 10\%$) by assuming that the emission region is composed of many cells, each containing a roughly uniform magnetic field that is randomly oriented from cell to cell (e.g. \citealt{Jones1985, Jones1988, Marscher2014, Kiehlmann2016}). 

Deterministic models usually consider the polarization due to a large scale helical magnetic field (e.g. \citealt{Lyutikov2005}), velocity shear (e.g. \citealt{Laing1980, DArcangelo2009}), or the compression of an initially tangled magnetic field by shock waves in the jet (shock-in-jet model, \citealt{Marscher1985, Hughes1985, Cawthorne1990}). For a helical magnetic field geometry, the net magnetic field seen by the observer can appear to be either transverse or longitudinal \citep{Lyutikov2005}, while an initially turbulent magnetic field can be partially ordered through velocity shear or shocks propagating in the jet. Velocity shear can arise when fast plasma near the jet axis flows past slower material closer to the boundary. The resulting velocity gradient stretches and aligns the magnetic field along the direction of flow, leading to transverse polarization angles (e.g. \citealt{Laing1980, DArcangelo2009}). Shocks can partially order a turbulent magnetic field by compressing the magnetic field component parallel to the shock front. For transverse or oblique shocks (e.g. \citealt{Hughes1985, Lister1998, Hughes2011}), the shock front is oriented transverse to the jet axis or at an oblique angle, resulting in polarization angles that are aligned with the jet axis\footnote{Relativistic aberration tends to make larger viewing angles appear smaller such that an oblique shock will appear to be oriented closer to the jet viewing angle, thereby appearing to be a transverse shock.}. For conical shocks, the polarization angle can be either parallel or perpendicular to the jet axis \citep{Cawthorne1990}. However, the largest possible polarization for the transverse case is $\sim 10$\%. 

Two component models, where the polarization is attributed to a variable component of higher polarization degree superposed on a constant component with lower polarization degree, have also been successful at modelling the observed polarization of blazars (e.g. \citealt{HagenThorn2008, Barres2010, Sorcia2013, Gaur2014, Bhatta2015}). The constant component is usually identified with persistent emission from the quiescent jet, while the variable component is attributed to a shock. For the shock-in-jet model, ordering of the magnetic field in the shocked region leads to a positive correlation between the polarization degree and total flux (e.g. \citealt{HagenThorn2008, Sorcia2013}) if the quiescent jet has a completely chaotic magnetic field. However, if the magnetic field in the unshocked region possesses a component parallel to the jet axis, the shock may strengthen the total flux but partially cancel the polarization, leading to an increase in total flux and a decrease in polarization degree (e.g. \citealt{HagenThorn2002, Gaur2014}). 

It is not clear which model best describes the photopolarimetric behaviour of flaring blazars. Here, we present quasi-simultaneous multiband photometric and polarization observations of \pks{} during a prominent optical flare in 2010, in an effort to understand the origin of the observed variability. Measurements were obtained from the long-term monitoring campaign of gamma-ray emitting blazars operated by the Steward Observatory and the SMARTS blazar programme. The polarized and photometric fluxes are compared and analyzed in terms of a two-component model. The paper is organized as follows: section~\ref{sec:obs} describes the observations, followed by a photopolarimetric analysis of the source in section~\ref{sec:analysis}, a discussion of the results is presented in section~\ref{sec:discuss} and the conclusions are put forth in section~\ref{sec:conclusions}. 

\section{Observations} \label{sec:obs}  
\pks{} is a target of the long-term monitoring of the optical polarization of Fermi detected gamma-ray blazars, operated by the Steward Observatory \citep{Smith2009}, and of the Small and Moderate Aperture Research Telescope System (SMARTS), a long-term photometric monitoring programme operated by the Cerro Tololo Inter-American Observatory \citep{Bonning2012}. Photopolarimetric measurements of the source were obtained from the publicly available SMARTS and Steward Observatory archives\footnote{The Steward Observatory archive can be found at \url{http://james.as.arizona.edu/~psmith/Fermi/}. The SMARTS archive can be found at \url{http://www.astro.yale.edu/smarts/glast/home.php}}. 

The Steward Observatory's polarization measurements are performed using SPOL, an imaging spectro-polarimeter consisting of polarizing optics, a spectrograph and a CCD imaging camera. The instrument is mounted on the 1.54 m Kuiper telescope or the 2.3 m Bok telescope of the Steward observatory. Light incident on the telescope is passed through an achromatic $\frac{1}{4}$-- and $\frac{1}{2}$--wave plate, which measures circular and linear polarization, respectively. A Wollaston prism separates the incident beam into its two orthogonal components, thereby enabling two independent measurements of the polarization. The polarization in the \Bf--, \Vf-- and \Rf--band were calculated by multiplying the relative Stokes spectrum (measured between 4355 \AA{} -- 7195 \AA) with the \Bf--, \Vf-- and \Rf--band spectral response functions of the telescope.

The SMARTS photometric observations are performed at optical and near-infrared wavelengths using four small telescopes (1.5m, 1.3m, 1.0m and 0.9m) that are located in Chile. Measurements are obtained with ANDICAM, a dual-channel imager with a dichroic that feeds an optical CCD and an infrared (IR) imager, resulting in  simultaneous $B$--, $V$--, $R$--, $J$-- and $K$--band observations. Photometric errors as low as 0.01 mag in the optical and 0.02 mag in the IR are observed for bright sources (<16 mag in the optical and <13 mag in IR).

\begin{figure}
  \centering
	\includegraphics[trim = 1.25cm 2.cm 1.4cm .54cm, clip, width=\columnwidth]{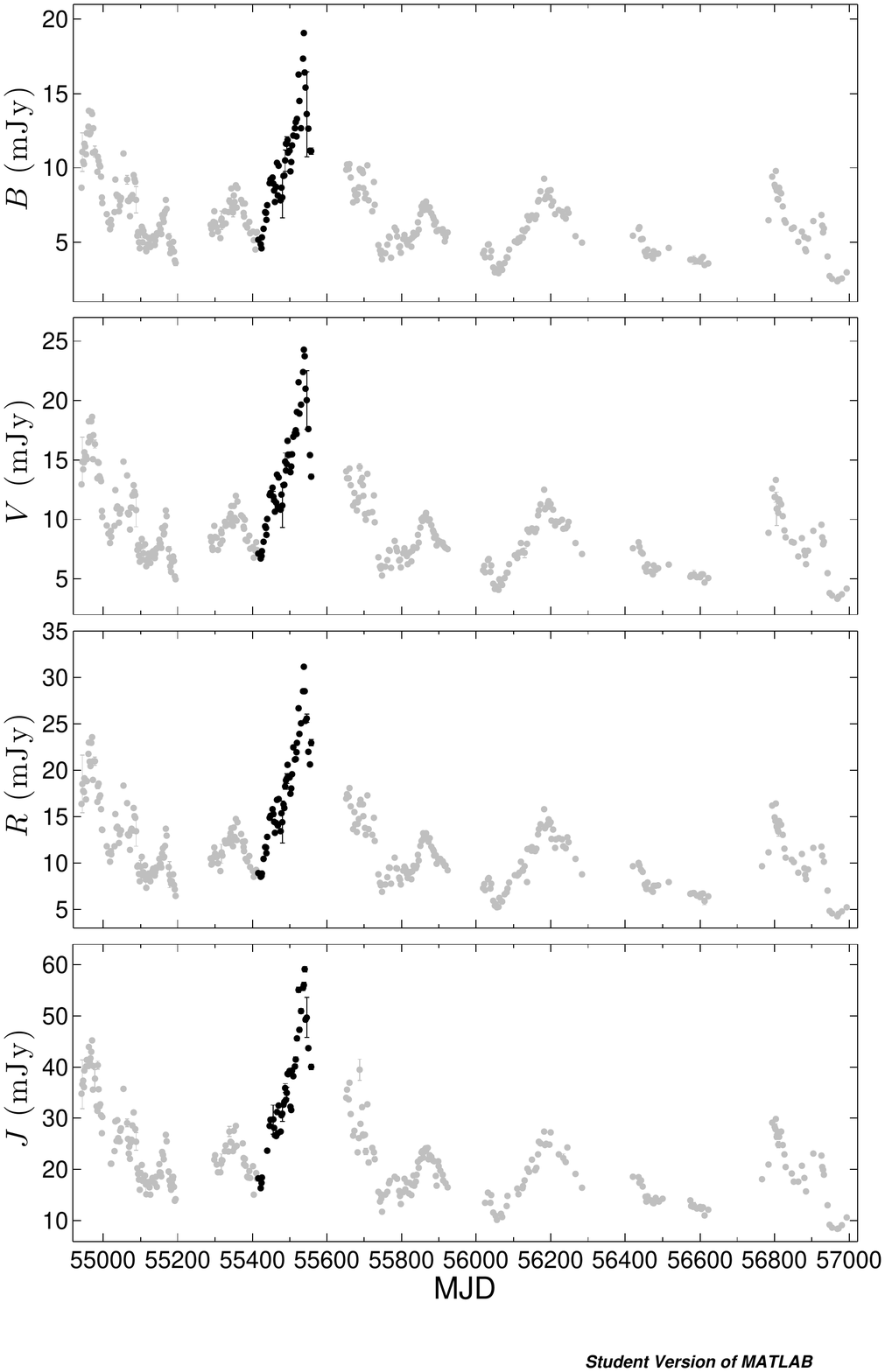}
  \caption{Extinction-corrected multiband light curves of \pks{} between April 2009 and Dec 2014. The most prominent flare occurred in 2010 and is shown in black. The flare is visible for all of the observed bands.}
  \label{fig:fluxobs}
\end{figure}

\begin{figure}
  \centering
	\includegraphics[trim = .86cm 2.05cm 1.57cm .55cm, clip, width=\columnwidth]{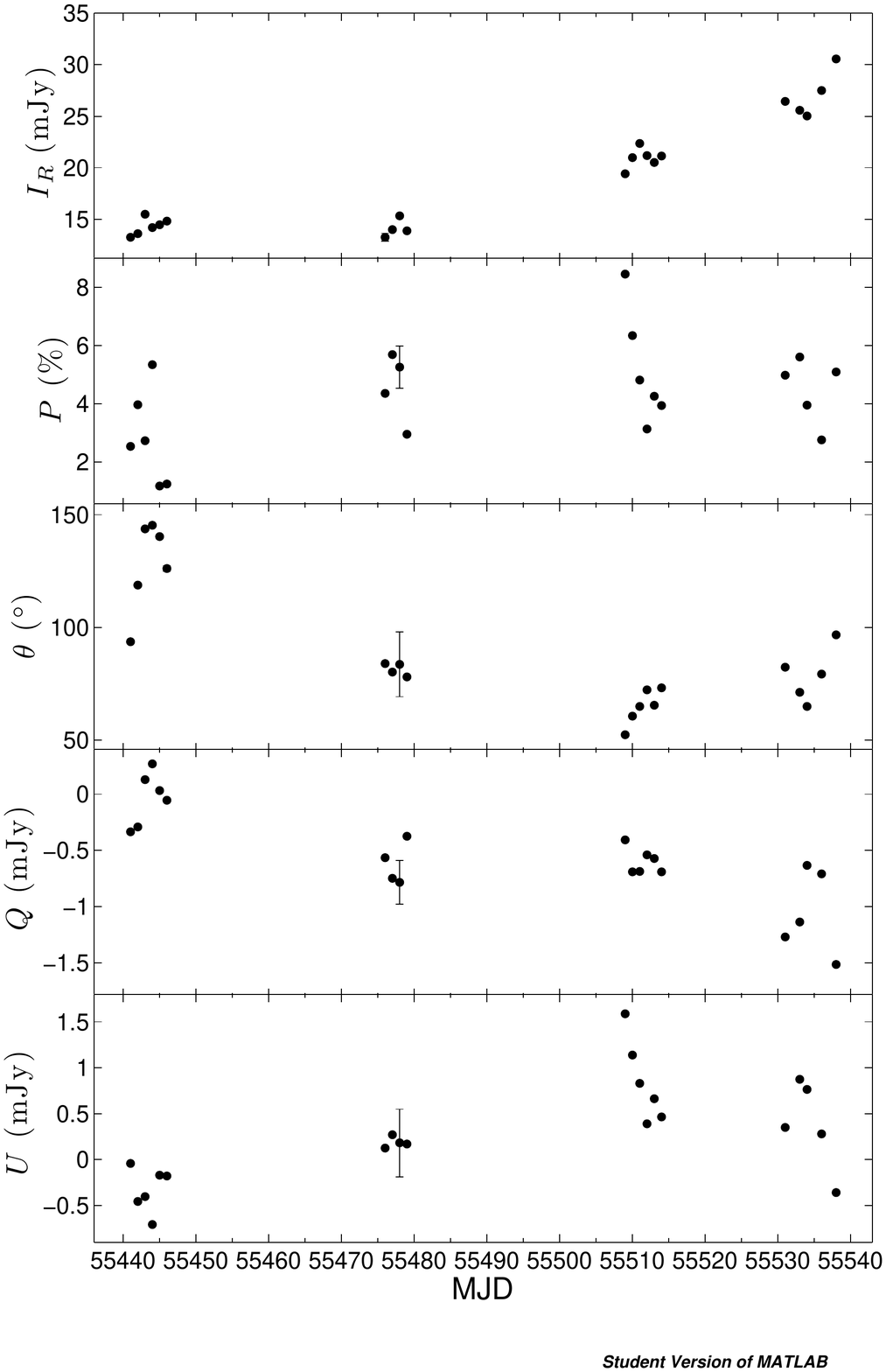}
  \caption{Simultaneous photopolarimetric observations of \pks{} during its 2010 optical outburst. The panels from top to bottom display the photometric flux,  polarization degree, polarization angle, absolute Stokes $Q$ flux and the absolute Stokes $U$ flux. All measurements were obtained in the \Rf--band.}
  \label{fig:mwlobs}
\end{figure}

The results of the photometric observations between April 2009 and Dec 2014 are shown in Fig.~\ref{fig:fluxobs}. The light curves reveal the presence of multiple flares throughout the monitoring period. The most prominent optical outburst occurs in 2010 (visible across all bands), with the source reaching a peak flux of $I_R = 30.5\pm 1.7$ mJy in the \Rf--band on $\mjd=55538$. Figure ~\ref{fig:mwlobs} displays simultaneous photopolarimetric measurements of \pks{} for the flare. The figure demonstrates erratic variability for the polarization degree, while the polarization angle or electric vector position angle (EVPA) appears to follow an overall decreasing trend, changing by approximately 90\dgr. At the onset of the flare, the polarization angle is aligned with the jet axis ($\theta = 150^\circ-160^\circ$, \citealt{Piner2008, Piner2010}), thereafter changing to an EVPA that lies roughly transverse to the jet direction as the flare grows.

\section{PhotoPolarimetric Analysis}\label{sec:analysis}
\begin{figure*}
  \centering
	\includegraphics[trim = 3.5cm 2.2cm 3.3cm 2.2cm, clip, width=1.25\columnwidth]{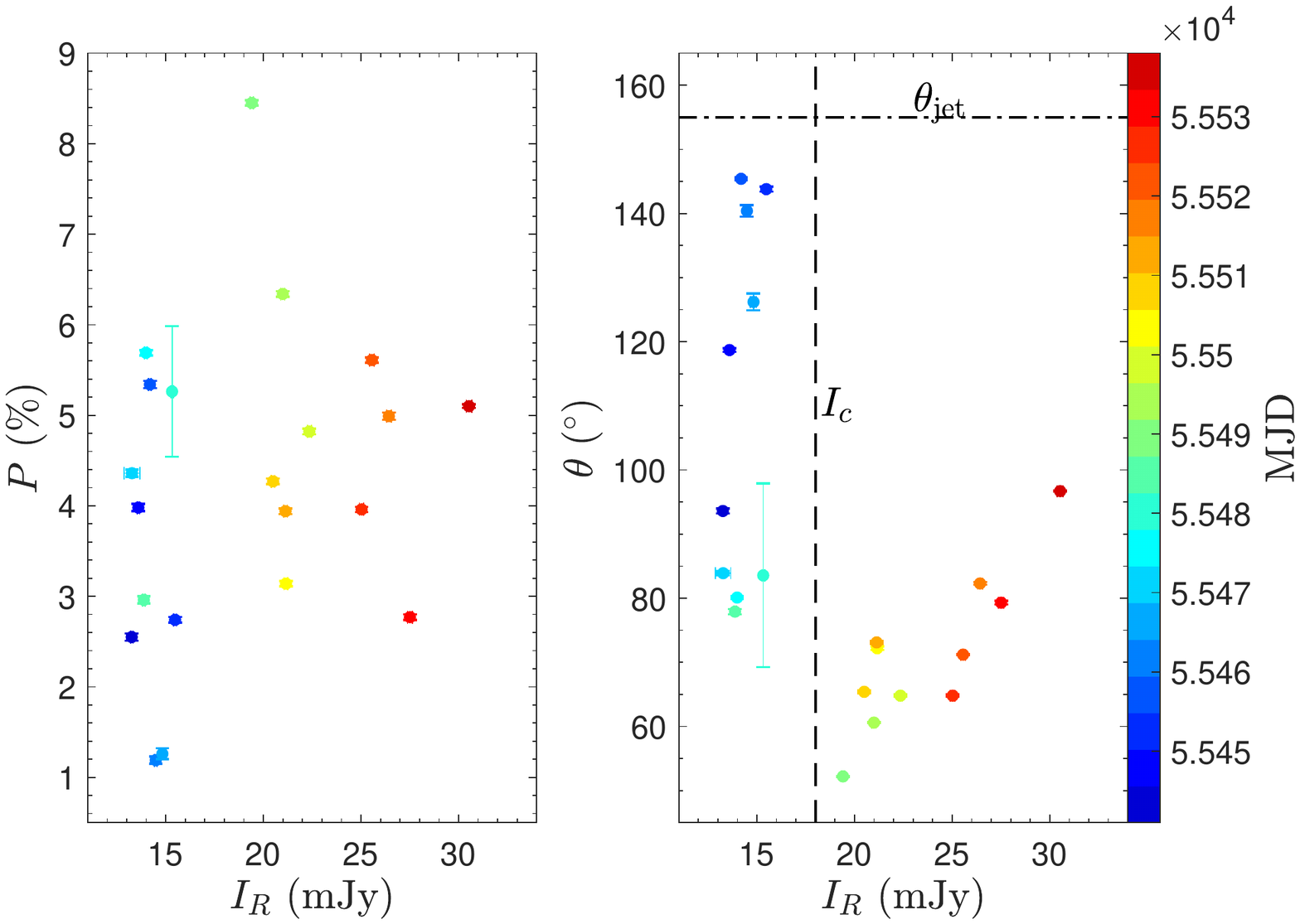}
  \caption{\textit{Left:} Dependence between the polarization degree and \Rf--band flux. \textit{Right:} Dependence between polarization angle and \Rf-band flux. Note two distinct states above and below $I_c=18$ mJy (dashed line). The dotted-dashed line represents the jet direction. Colour indicates the date of the observation, with the scale given by the colour bar.}
  \label{fig:polflux}
\end{figure*}

Inspection of the relationship between the polarization angle and \Rf--band flux show the existence of two distinct states at low and high fluxes (see Fig.~\ref{fig:polflux}). Below 18 mJy, the polarization angle takes on a range of values ($\sim75^\circ-150$\dgr) without any clear relation to the flux, while a positive correlation is observed between the EVPA and \Rf--band flux above 18 mJy (with a correlation coefficient $r=0.84$), with the polarization angle oriented roughly transverse to the jet direction ($60^\circ-70^\circ$) for the epoch of highest polarization. The polarization degree, in contrast, does not demonstrate a clear correlation with the flux. Although the maximum polarization degree is detected during the high flux state. 

\subsection{Polarization Parameters of the Variable Component}
Following \cite{HagenThorn1999}, the observed emission is decomposed as the superposition of a variable component and a constant component, $I = I_\var+I_\cons$, which yields:
\begin{equation}\label{eq:varpolflux}
\displaystyle
\begin{cases}
Q = q_\var I + (Q_\cons-q_\var I_\cons)\\
U = u_\var I + (U_\cons-u_\var I_\cons),
\end{cases}
\end{equation} 
where the subscripts $\var$ and $\cons$ denote the variable and constant emission component, respectively. Then, if the variable component has constant polarization properties (i.e. $q_\var$ and $u_\var$ is constant), a linear relationship will be observed for \textit{Q} versus \textit{I} and \textit{U} versus \textit{I}. The slopes of these lines are the relative Stokes parameters of the variable component (see equation~\ref{eq:varpolflux}), which give the polarization degree and EVPA of the variable emission component. Hence, a linear relation in the space of the Stokes parameters $\{I,Q,U\}$ suggests that the observed emission is due to a single variable component with constant polarization degree and angle. 

\begin{figure*}
  \centering
	\includegraphics[trim = 2.8cm 2.2cm 3.2cm 2.2cm, clip, width=1.25\columnwidth]{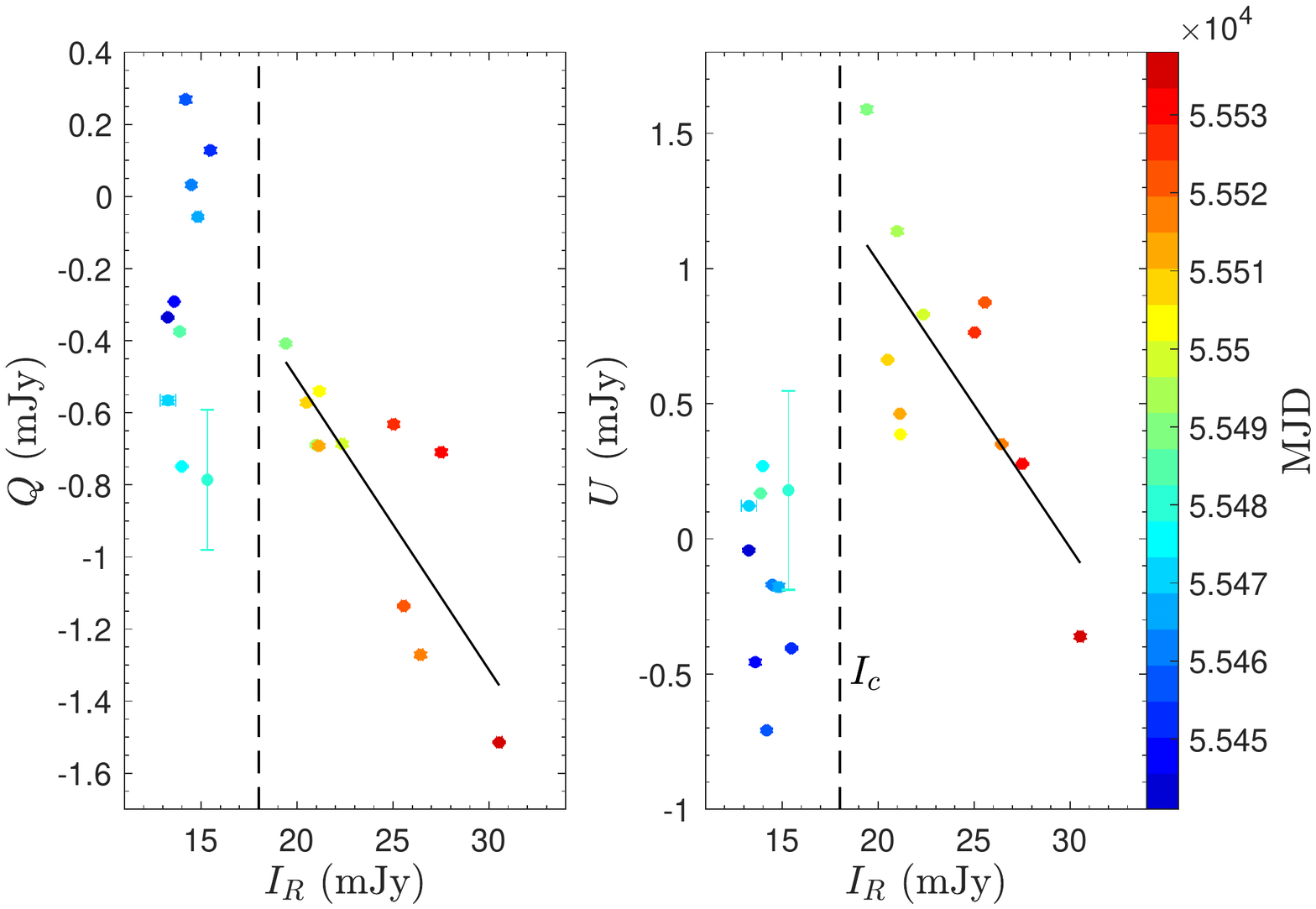}
  \caption{Comparison of the absolute Stokes parameters during the optical flare seen in 2010. The linear fit to the data for all $I>18$ mJy is indicated by the solid line. The dashed line represents $I_c=18$ mJy. Colour indicates the date of the observation, with the scale given by the colour bar.}
  \label{fig:stokes}
\end{figure*}

Figure~\ref{fig:stokes} displays the relationship between the Stokes fluxes and the \Rf--band flux. Inspection of Fig.~\ref{fig:stokes} reveals two distinct types of behaviour. The Stokes $Q$ and $U$ flux appear to vary randomly between $-0.8$ mJy and 0.3 mJy below 18 mJy, while there is a linear relationship between the variables above 18 mJy with a correlation coefficient $r_{Q-I}=-0.82$ for $Q$ vs $I$ and $r_{U-I}=-0.72$ for $U$ vs $I$. The best fit line, calculated using the orthogonal regression method, is superimposed. The slope of each line gives the relative Stokes parameters of the variable component which yields $p_\mathrm{var}=13.3\pm 2.8$\% for the polarization degree and $\theta_\mathrm{var}=116\pm 6$\dgr{} for the polarization angle of the variable component. The polarization degree is comparable to the highest fractional polarization for the variable component found by \cite{Barres2010}, $p=12.5\%$. The derived polarization degree is relatively small when compared to the maximum possible polarization for a \sy{} source with $\alpha=1.12$ (cf. section~\ref{subsect:sed}) in a uniform magnetic field ($p=76\%$), which is indicative of a non-uniform magnetic field direction inside the emitting region.

Below 18 mJy, the polarization variability appears to be well represented by a uniform distribution. Random sampling $Q$ and $U$ from the uniform distribution on the same interval $[-0.8, 0.3]$ yields a variance $\sigma^2_{Q_m}=0.13$ and $\sigma^2_{U_m}=0.15$, which is comparable to the observed values, $\sigma^2_Q=0.13$ and $\sigma^2_U=0.10$. Therefore, the polarization variability is well-represented by a uniform distribution during the low flux state.

\subsection{Spectral Energy Distribution of the Variable Component}\label{subsect:sed}
For photometry, the flux ratio for different pairs of spectral bands are \citep{HagenThorn1999}:
\begin{equation}
\lf\frac{I_\nu}{I_{\nu_0}}\rt^\var=\upalpha_{\nu\nu_0},
\end{equation}
where the superscript refers to the variable emission component and $\nu_0$ is the reference spectral band. If the spectral index of the variable emission does not change with frequency then these flux ratios are constant and 
\begin{equation}\label{eq:varflux}
I_\nu = \upalpha_{\nu\nu_0} I_{\nu_0}+\lf I_{\nu}^\cons-\upalpha_{\nu\nu_0} I_{\nu_0}^\cons\rt.
\end{equation}
Equation~\ref{eq:varflux} shows that a linear relationship will be observed for $I_\nu$ versus $I_{\nu_0}$ for multicolour observations. Hence, a linear relation in the flux-flux diagrams for several bands indicates that the relative SED of the variable component remains steady. The spectral index of the variable emission can then be derived from the slope of the best fit line for $\log \upalpha_{\nu\nu_0}$ versus $\log\nu$.

\begin{figure}
  \centering
	\includegraphics[trim = 1.85cm 6.65cm 1.6cm 9.2cm, clip, width=0.95\columnwidth]{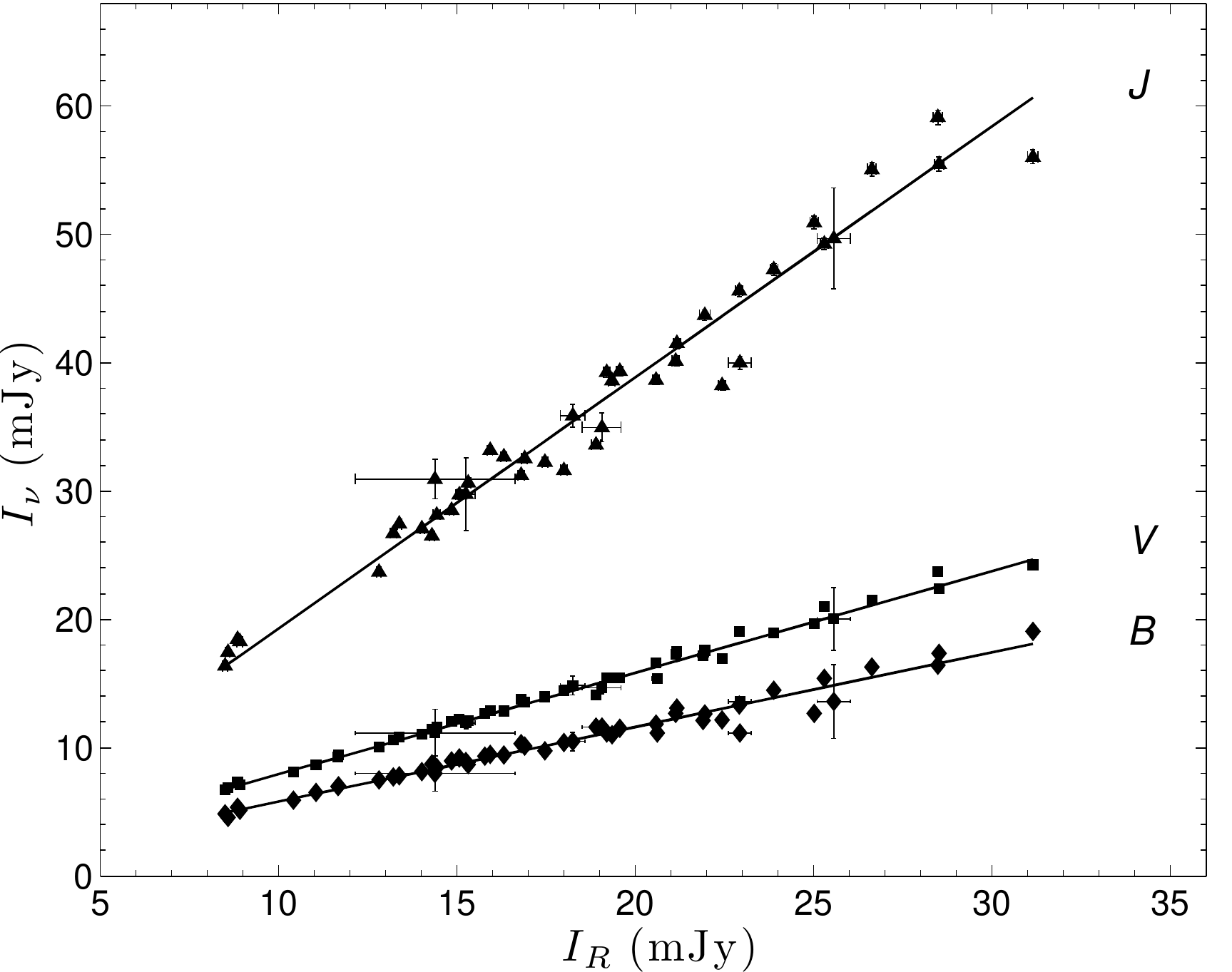}
  \caption{The \Bf--, \Vf-- and \Jf--band fluxes relative to the \Rf--band flux during the 2010 optical flare of \pks. The best fit lines are superimposed.}
  \label{fig:fluxflux}
\end{figure}

\begin{figure}
  \centering
	\includegraphics[trim = 1.7cm 6.65cm 1.9cm 9.2cm, clip, width=0.95\columnwidth]{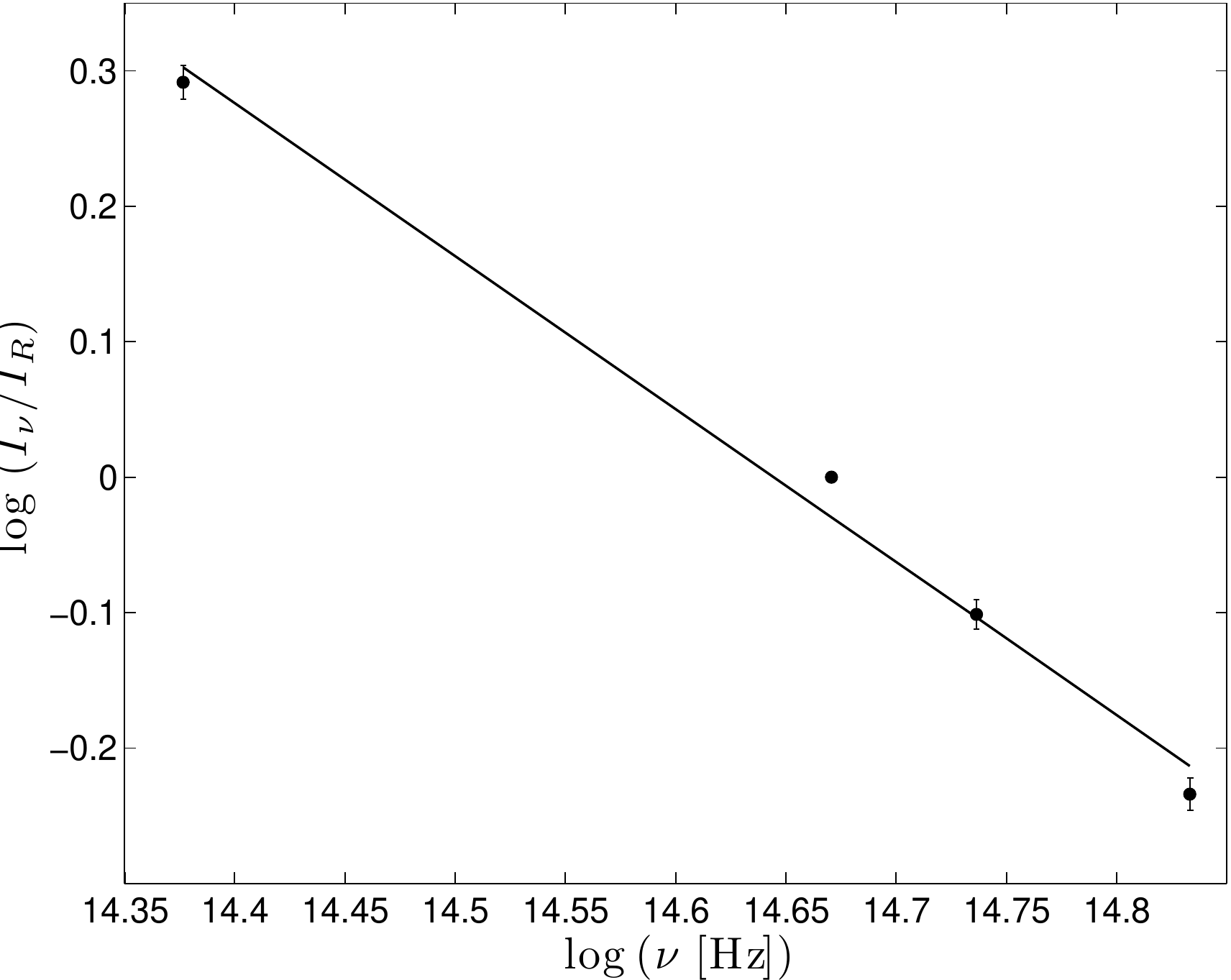}
  \caption{Relative SED of the variable emission component of \pks{} during its 2010 outburst. The best fit line is superimposed, with slope $\alpha=1.12\pm 0.07$.}
  \label{fig:sed}
\end{figure}

The \Bf--, \Vf-- and \Jf--band fluxes relative to the \Rf--band flux is displayed in Fig.~\ref{fig:fluxflux}, with the best fit lines calculated using the orthogonal regression method. The slopes of the fitted lines represent the flux ratio between the given pairs of bands. Since a linear relationship is observed it can be inferred that the relative SED of the variable component remains steady during the observation period. The variable emission spectrum of the source is displayed in Fig.~\ref{fig:sed}, which shows that the SED is well-described by a power-law $F_\nu\propto \nu^{-\alpha}$ with slope $\alpha=1.12\pm 0.07$, consistent with emission from a \sy{} source. 

\subsection{Variability Timescale}
Following \cite{Burbidge1974}, the timescale of variability for the multiband light curves is defined as $\tau=dt/\ln(I_1/I_2)$, where $dt$ is the time interval between flux measurements $I_1$ and $I_2$, with $I_1>I_2$. All possible timescales $\tau_{ij}$ are calculated for any pair of observations for which $|I_i-I_j|>\sigma_{I_i}+\sigma_{I_j}$. The minimum variability timescale is then given by $\tau=\text{min}\{\tau_{ij}\}$, where $i=1...N-1$, $j=i+1,...N$ and $N$ is the number of observations. The results are listed in Table~\ref{tab:vartime}. The columns are (1) the observation band, with $I_P$ the polarized flux in the \Rf--band, (2) the number of observations $N$ and (3) the variability timescale $\tau$ (days). Table~\ref{tab:vartime} indicates that the timescale of variability during the flare is on the order of a few days and that light curves with similar sampling display an increase of timescale with wavelength. 

The shortest timescale, $\tau\sim 0.7$ days, occurred in the polarized flux. Comparison of the variability timescales of the total and polarized \Rf--band fluxes implies that the polarized emission originates in a subsection ($\sim \frac{1}{16}$) of the optical emission region. 
\begin{table}
	\centering
	\caption{Variability timescales of \pks{} during its 2010 outburst} \label{tab:vartime}
	\begin{tabular}{ccc} 
		\hline
		(1)	&(2)	&(3)\\	
		Band  &$N$	&$\tau_\mathrm{min}$\\
		    &    &(days)\\
		\hline
		$B$		&49		&8.38\\
    	$V$		&49			&9.09\\
    	$R$		&49		&10.75\\
    	$J$		&42			&9.16\\
    	$I_P$	&21			&0.68 \\
		\hline
	\end{tabular}
\end{table}

\section{Discussion} \label{sec:discuss}
Simultaneous polarimetric and photometric observations of \pks{} show that the optical flare seen in 2010 can be attributed to a single variable component with the following properties:
\begin{enumerate}[label=\arabic*.]
\item a steady spectral shape,
\item a high degree of polarization,
\item a correlation between the flux and polarization angle,
\item a tendency of the magnetic field to be transverse to the jet direction during the epoch of highest polarization, and 
\item an increase in the timescale of variability with wavelength.
\end{enumerate}
These characteristics are consistent with a shock propagating in a relativistic jet with a turbulent magnetic field \citep{Marscher1985, Hughes1985}, with \sy{} radiation losses leading to larger variability timescales at longer wavelengths. The shock orders the turbulent magnetic field along the shock front, leading to a change in the polarization angle. Since transverse shocks lead to polarization angles oriented parallel to the jet axis, it is more likely that an oblique shock is responsible for the observed emission. 

\subsection{Shock-in-Jet Model}
The observed flux of a shock moving through a turbulent plasma with constant bulk Lorentz factor $\Gamma$ is:
\begin{equation}\label{eq:shockflux}
F=F_0\nu^{-\alpha}\delta^{(3+\alpha)}\delta'^{(2+\alpha)},
\end{equation}
where $F_0$ is the flux scaling factor, $\delta=[\Gamma(1-\beta\cos\Phi)]^{-1}$ is the Doppler factor of the jet in the observer's frame, $\beta=\sqrt{1-\Gamma^{-2}}$ is the speed of the shock normalized to the speed of light and $\Phi$ is the viewing angle of the jet in the observer's frame. The factor $\delta'$ is the Doppler factor of the shocked plasma in the rest frame of the shock. Without loss of generality, it can be assumed that the velocity of the shocked plasma in the  frame of the shock is $\ll c$, where $c$ is the speed of light and $\delta'\approx 1$ \citep{HagenThorn2008}.

\begin{figure}
  \centering
	\includegraphics[trim = 1.04cm 2.cm 1.2cm 0.68cm, clip, width=\columnwidth]{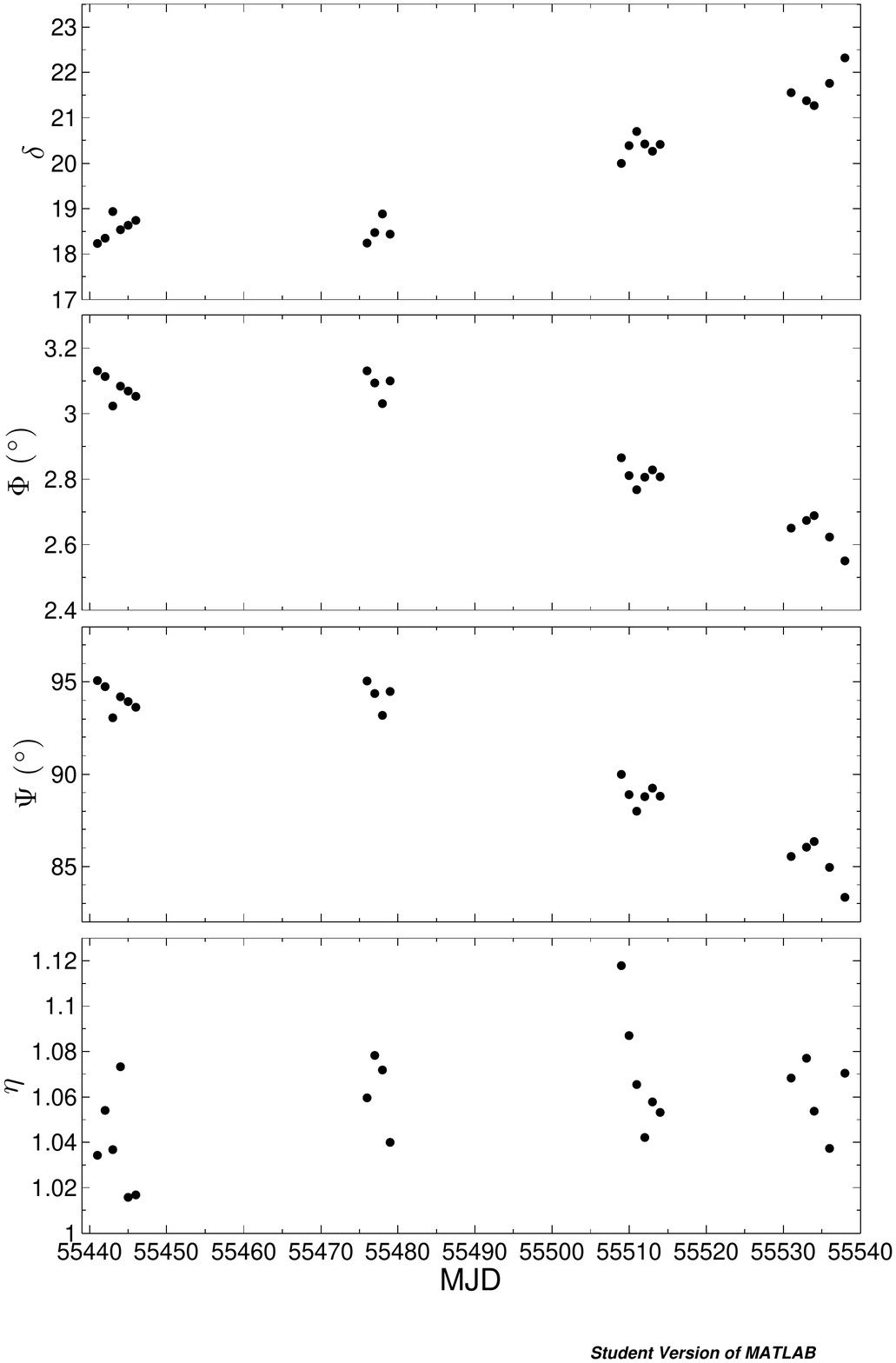}
  \caption{Derived values for the Doppler factor $\delta$, jet viewing angle $\Phi$, viewing angle of the shock $\Psi$ and compression factor of the shocked plasma $\eta$ for \pks{} during its 2010 optical outburst.}
  \label{fig:shock}
\end{figure}

From \cite{Hughes1991}, the polarization degree of the shocked plasma is:
\begin{equation}\label{eq:pol}
p\approx \frac{\alpha+1}{\alpha+5/3}\frac{(1-\eta^{-2})\sin^2\Psi}{2-(1-\eta^{-2})\sin^2\Psi},
\end{equation}
where $\eta=n_\mathrm{shock}/n_\mathrm{unshocked}$ is the density of the shocked region relative to the unshocked region and $\Psi$ is the viewing angle of the shock in the observer's frame, which is subject to relativistic aberration and is defined as: 
\begin{equation}\label{eq:viewang}
\Psi=\tan^{-1}\left\{\sin\Phi/\left[\Gamma\left(\cos\Phi-\sqrt{1-\Gamma^{-2}}\right)\right]\right\}
\end{equation}
Typical bulk Lorentz factors derived from emission models of the VHE emission of \pks{} during the high state seen in 2006 yield $\Gamma=25-50$ \citep{Foschini2007, Begelman2008, Narayan2012}, while \cite{Reynoso2012} showed that the observed VHE emission could also be explained by multiple shocks with $\Gamma = 10.5-13$. The Doppler factor $\delta(t)$ can then be determined from equation~\ref{eq:shockflux} by adopting $\alpha=1.12$, the spectral index of the variable component, and an average value of $\Gamma\sim 20$. The scaling factor $F_0=F_\mathrm{max}\nu^\alpha/\delta_0^{(3+\alpha)}$, with $F_\mathrm{max}$ the observed flux in the \Rf\dash band at peak polarization. The $\delta_0$ factor is calculated from $\Phi_0=2.9$\dgr, which is determined from equation~\ref{eq:viewang} for $\Psi=90$\dgr, corresponding to the maximum polarization degree due to the shock. 
The evolution of the shock parameters are found by using $\delta(t)$ to solve for $\Phi(t)$, which is then used to solve for the viewing angle of the shock, $\Psi(t)$, through equation~\ref{eq:viewang}. The shock compression factor, $\eta(t)$, can then be derived by substituting $\Psi(t)$ and the observed polarization degree into equation~\ref{eq:pol}. The resulting shock parameters are displayed in Fig.~\ref{fig:shock} and demonstrate that, for constant bulk Lorentz factor, small changes in the viewing angle ($<1^\circ$) and plasma compression factor ($\Delta \eta = 0.10$) produce large variations in the flux and polarization degree. The peak flux is reached when the viewing angle of the jet is a minimum ($\Phi=2.6$\dgr). The plasma compression reaches its highest value during the rising phase of the flare, roughly 30 days before the total brightness reaches its peak value. However, $\eta$ does not appear to have any systematic trends. 

\begin{figure*}
  \centering
	\includegraphics[trim = 0cm 0cm 0cm 0cm, clip, angle = 0, width=1.5\columnwidth]{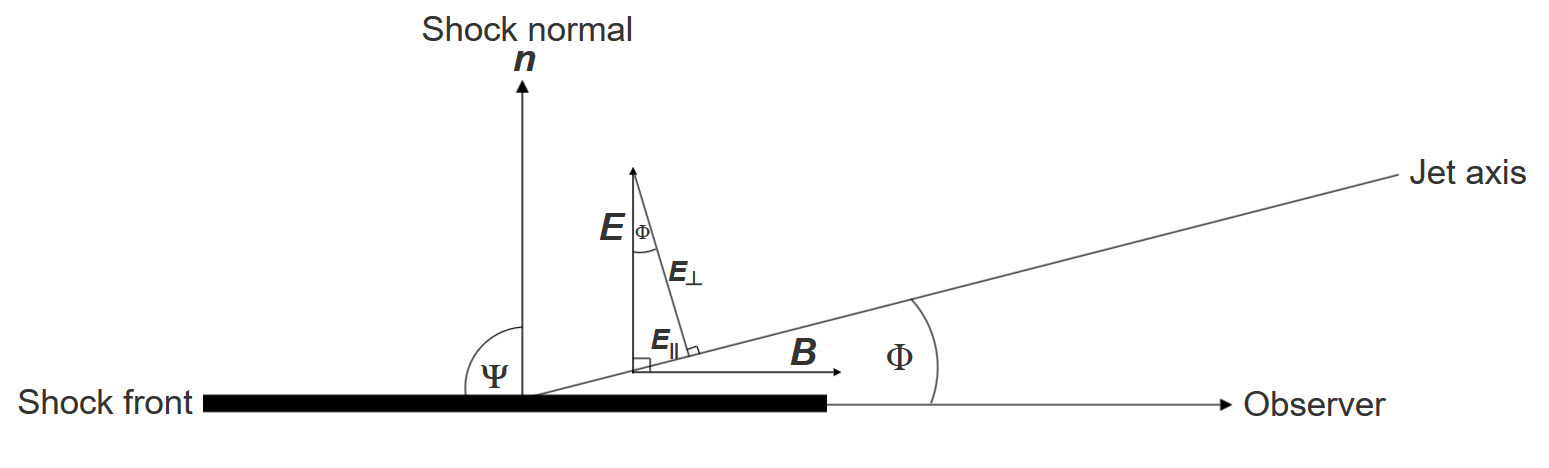}
  \caption{Electric field arising from shock compression of a tangled magnetic field for an edge-on shock ($\Psi = 90$\dgr) for a blazar with viewing angle $\Phi$.}
  \label{fig:shockgeometry}
\end{figure*}

The shock is seen nearly edge-on throughout the flaring period ($\Psi\sim 91$\dgr), varying by $<12$\dgr, which implies that the shock is seen at an oblique angle to the jet axis (see Fig.~\ref{fig:shockgeometry}). Compression of the magnetic field by the shock yields a net magnetic field component parallel to the shock front. The parallel and perpendicular magnetic field components with respect to the jet axis are then given by $B_\parallel = B\cos\Phi$ and $B_\perp = B\sin\Phi$, respectively. The corresponding electric field components perpendicular and parallel to the jet axis are $E_\perp = E\cos\Phi$ and $E_\parallel = E\sin\Phi$ (see Fig.~\ref{fig:shockgeometry}). For the small angle approximation relevant for blazars the electric field components reduce to $E_\perp\approx E$ and $E_\parallel\approx 0$. Hence, the polarization angle for an edge-on shock is expected to lie transverse to the jet axis, as demonstrated in Fig.~\ref{fig:mwlobs}, which shows that the polarization angle lies roughly transverse to the jet axis during the peak polarization epoch ($\theta\sim 60^\circ-70$\dgr). 

For the shock-in-jet scenario, relativistic electrons are injected into the emitting region by the shock, leading to a steady spectral shape and short timescale of the variability in the optical bands. A steady spectral shape implies a quasi-steady state of the emission between the rate of injection and radiative losses, which is re-established every $8-11$ lt-days. The timescale of variability relates to the thickness of the shock front, which is determined by the lifetime of the relativistic electrons being accelerated at the front. The lifetime of the \sy{} electrons for a given frequency $\nu$ in GHz in the observer's frame is \cite{HagenThorn2008}:
\begin{equation}\label{eq:synclife}
t_\mathrm{sync}=4.75\times 10^2\left(\frac{1+z}{\delta\nu_\mathrm{GHz}B^3_\mathrm{G}}\right)^{1/2}\; \mathrm{days,}
\end{equation}  
where $B_\mathrm{G}$ is the magnetic field in gauss. Adopting the maximum derived Doppler factor for the observations ($\delta=22.3$) and $\tau=10.75$ days in the \Rf--band yields $B=0.06$ G, in agreement with other estimates of the typical magnetic field in blazars (e.g., \citealt{Marscher1985, HagenThorn2008, Barres2010, Sorcia2013}). Equation \ref{eq:synclife} also implies that the timescale of variability in the \Rf--band should be a factor of 1.2 greater than the \Bf--band, which is consistent with what is observed (see Table~\ref{tab:vartime}). Therefore, the polarimetric and multicolor photometric behaviour of \pks{} during its 2010 outburst is consistent with the characteristics of the shock-in-jet model. 

\subsection{Helical Magnetic Field}
An alternative interpretation for the photopolarimetric behaviour of \pks{} during its 2010 optical flare is that the observed variations are due to changes in the bulk Lorentz factor and viewing angle of a jet with a helical magnetic field geometry.

\begin{figure}
  \centering
	\includegraphics[trim = 1.83cm 6.9cm 0.98cm 9.1cm, clip, width=\columnwidth]{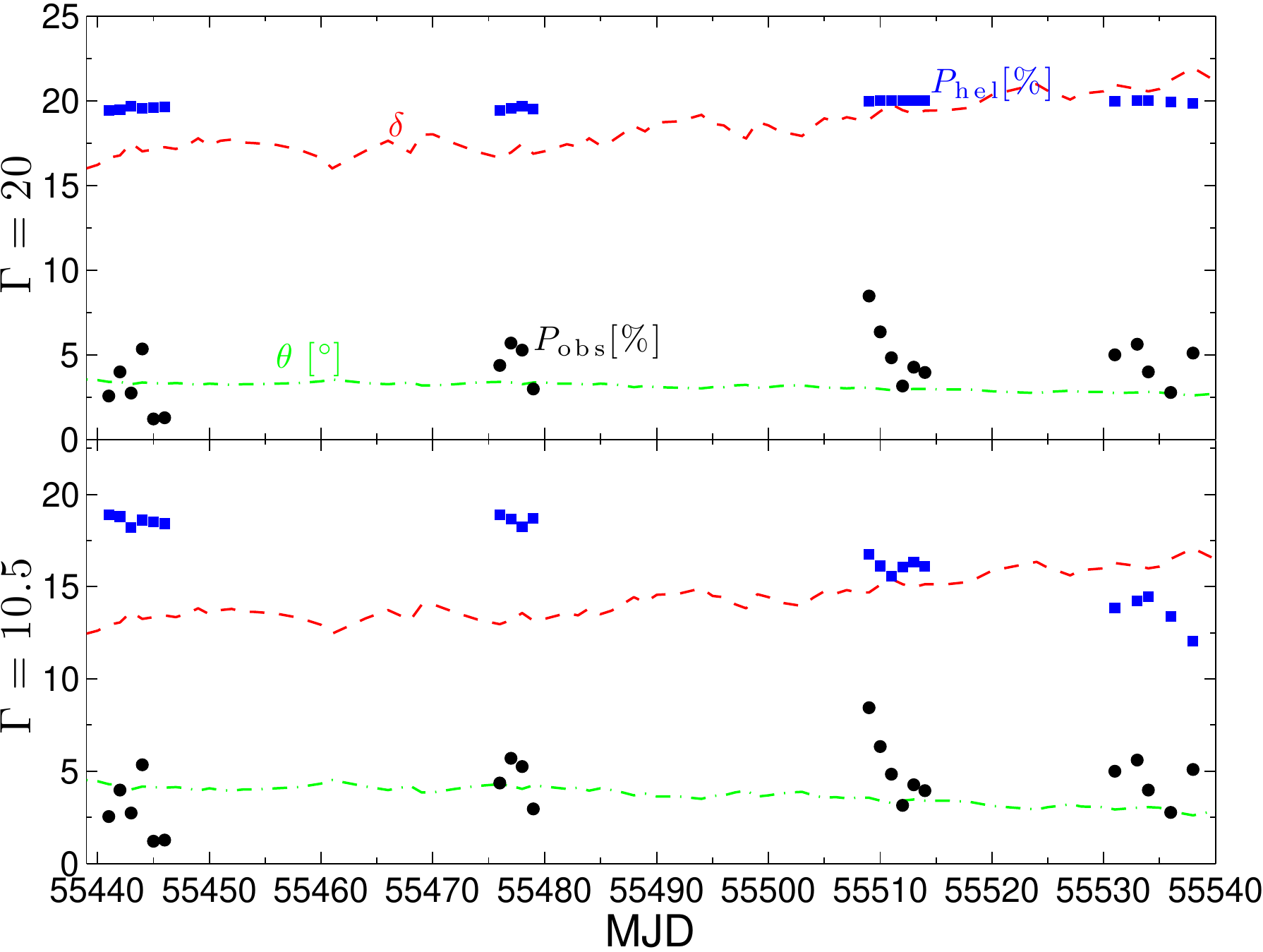}
  \caption{Top panel: The Doppler factor $\delta$ (red dashed line) and viewing angle $\theta$ (green dotted-dashed line) characterizing the optical emission region according to a geometrical interpretation of the long-term optical flux variability for a bulk Lorentz factor $\Gamma=20$. The observed polarization degree is represented by the black filled circles, while the polarization degree as predicted by the helical magnetic field model is indicated by the blue filled circles. The $\Gamma=10.5$ case is shown in the bottom panel for comparison.  Models of the quiescent emission give $\Gamma=10$ \citep{Reynoso2012}.}
  \label{fig:helmod}
\end{figure}

For a jet pervaded by a helical magnetic field, the observed polarization for optically thin synchrotron emission\footnote{For the diffuse and reverse-field pinch cases of a filled jet, where the number density of relativistic particles is proportional to the square of the magnetic field (see fig. 11c and 12c in \cite{Lyutikov2005}).} can be approximated as \citep{Lyutikov2005}: 
\begin{equation}\label{eq:hel}
  P = P_\mathrm{max}\sin^2\phi,
\end{equation}
with $P_\mathrm{max}\approx 20\%$ and $\phi$ the viewing angle in the rest frame of the jet, which is related to the observed angle $\Phi$ through the Lorentz transformation:
\begin{equation}\label{eq:abang}
  \sin\phi = \delta\sin\Phi,
\end{equation}
where $\delta$ is the Doppler factor of the jet in the observer's frame. The Doppler factor can be obtained from the observed flux due to a relativistic plasma with bulk Lorentz factor $\Gamma$ for a smooth, continuous jet:
\begin{equation}\label{eq:flux}
  F _\nu= \delta^{2+a}F'_\nu,
\end{equation}
where $F'_\nu\propto\nu'^{-\alpha}$ is the flux in the rest frame of the jet, $\nu'$ the frequency in the rest frame of the jet and $\alpha$ is the spectral index. Adopting $\alpha=1$ (comparable to value derived for the variable component) yields $\delta=\delta_\mathrm{max}(F/F_\mathrm{max})^{1/3}$, where $\delta_\mathrm{max}$ is determined from the definition of the Doppler factor for $\Phi_\mathrm{min}=2.6^\circ$ and $\Gamma=20$ and $F_\mathrm{max}$ is the maximum observed flux in the \Rf--band, and the viewing angle through the definition of the Doppler factor. The polarization degree for a helical magnetic field can then be recovered by substituting $\delta$ and $\Phi$ into equation~\ref{eq:abang} and ~\ref{eq:hel}. The results are displayed in Fig.~\ref{fig:helmod} and demonstrate that the helical magnetic field model overestimates the observed polarization degree. Adopting a lower value for the bulk Lorentz factor ($\Gamma=10.5$\footnote{Compare with models of the quiescent emission, which yield $\Gamma=10$ \citep{Reynoso2012}.} \citealt{Reynoso2012}.) yields a similar result, although the resulting polarization degree is closer to the observed level of polarization. Hence, it is unlikely that the observed polarization of \pks{} during its 2010 outburst is due to a helical magnetic field. 

\section{Conclusions}\label{sec:conclusions}
The BL Lac \pks{} experienced a prominent optical outburst in 2010, increasing by a factor of 3.7 over roughly 4 months. Analysis of multi-colour photometric measurements and \Rf--band polarization measurements indicate the following:
\begin{enumerate}[label=\arabic*.]
\item The existence of two distinct states at low and high fluxes. Below 18 mJy, the polarization angle and photometric flux is not correlated, while there is a positive correlation between the polarization angle and flux above 18 mJy ($r=0.84$).
\item During the high flux state, the polarization angle during the epoch of highest polarization tends to be oriented transverse to the jet direction.
\item The variable emission can be attributed to a variable component with high polarization degree (13.3\%) and a constant power-law spectral energy distribution with $\alpha=1.12\pm 0.07$.
\item Variations on timescales of days are present for all observation bands, with the minimum timescale of variability increasing with wavelength.
\end{enumerate}
These properties are consistent with the shock-in-jet model, which shows that the observed variability can be explained by an edge-on shock (assuming constant bulk Lorentz factor $\Gamma=20$). Some parameters derived for the relativistic jet within the shock-in-jet model are $B=0.06$ G for the magnetic field, $\delta=22.3$ for the Doppler factor and $\Phi=2.6^\circ$ for the viewing angle.

\section*{Acknowledgements}
We gratefully acknowledge the support of the South African National Research Foundation (NRF) and the Department of Science and Technology Department through the South African Square Kilometre Array project. Data from the Steward Observatory's spectropolarimetric monitoring project were used, which is supported by the Fermi Guest Investigator grants NNX08AW56G, NNX09AU10G, NNX12AO93G, and NNX15AU81G. This paper also makes use of SMARTS optical and near-infrared light curves. Special thanks to the referee for her, or his helpful suggestions. 




\bibliographystyle{mnras}
\bibliography{biblio} 







\bsp	
\label{lastpage}
\end{document}